\documentstyle[preprint,aps,prabib]{revtex}
\begin{document}
\draft
\title{Glueball Spin}
\author{D. Singleton
\thanks{E-mail address : dougs@csufresno.edu}}
\address{Dept. of Physics, CSU Fresno, 2345 East San Ramon Ave.
M/S 37, Fresno, CA 93740-8031}
\date{\today}
\maketitle
\begin{abstract}
The spin of a glueball is usually
taken as coming from the spin (and possibly the orbital
angular momentum) of its constituent gluons.
In light of the difficulties
in accounting for the spin of the proton from its
constituent quarks, the spin of glueballs is
reexamined. The starting point is the fundamental QCD
field angular momentum operator written in terms of the
chromoelectric and chromomagnetic fields. First, we look
at the possible restrictions placed on the structure
of glueballs from the requirement that the
QCD field angular momentum operator should satisfy
the standard commutation relationships.
This analysis can be compared to the electromagnetic
charge/monopole system, where the requirement that
the total field angular momentum obey the angular
momentum commutation relationships places restrictions
({\it i.e.} the Dirac condition) on the system.
Second, we look at the expectation value of the field
angular momentum operator under some simplifying
assumptions.
\end{abstract}
\tightenlines

\section{Introduction}

One of the results of the EMC experiment \cite{ashman}
was to show that the spin of the proton did not come
solely from the spin of the valence quarks
as naive quark models indicated. After the
EMC experiment other possible contributing sources
to the proton's spin were considered, such as
angular momentum from sea quarks,
angular momentum from the orbital motion of valence
quarks, and angular momentum from the chromodynamic field.

In light of this ``surprise'' \footnote{In Ref. \cite{sehgal}
there was already an early hint that the spin
of the proton might not be as simple as quark models
indicated} in the make up of the spin of the proton,
we want to take a closer look at the spin of pure glue
bound states. In the case of glueballs there is currently no
firm experimental evidence for their existence.
Ref. \cite{rpp} gives the
status of various candidates. Thus, presently there
is no need to explain anything in regards
to the spin of the glueballs, since there is no clear
cut discrepancy between experiment and the theoretical
models. However, since the simple quark models
did not work in accounting for the spin structure of
the proton, it is reasonable to ask if
a similar problem may exist for the simple
theoretical models of the glueball's spin. While most
glueball studies focus on the mass spectrum of glueballs,
here we want to examine the glueball's spin, starting
from the fundamental gluonic field angular momentum
operator given below in Eq. (\ref{1}).

The gauge-invariant QCD field angular
momentum operator \cite{ji} which is responsible for
the glueball's spin is
\begin{equation}
\label{1}
{\bf J} _{GB} = \int d^3 x \left[
{\bf x} \times ({\bf E}^a \times {\bf B}^a ) \right]
\end{equation}
where ${\bf E}^a , {\bf B}^a$ are the chromoelectric
and chromomagnetic fields, and $a$ is an SU(3) color index.
Except for the color indices this form of the
pure gauge field angular momentum is similar to
the Abelian case \cite{jackson}, and like the Abelian
form contains both spin and orbital contributions.
It is not possible to split Eq. (1) into a separate
spin and orbital part {\it in a gauge-invariant way}
\cite{ji1} \cite{jaffe}. Eq. (\ref{1})
is the total angular momentum of the
glueball, and so it should satisfy the
angular momentum commutation relationship
\begin{equation}
\label{2}
[J_{GB} ^i , J_{GB} ^j] = i \epsilon ^{ijk} J_{GB} ^k
\end{equation}
The main point of this paper is that the
requirement of Eq. (\ref{2}) will place
some restrictions on the form of
the chromodynamic fields, ${\bf E}^a , {\bf B}^a$,
and on the structure of glueballs.

Before proceeding with the details of the
calculation of $[J_{GB} ^i , J_{GB} ^j]$ we should ask
if it is physically reasonable that Eq. (\ref{2})
should place restrictions on the color
fields. For the electromagnetic interaction there are
also systems which carry field angular momentum,
such as the charge/monopole configuration. For this example
the angular momentum commutators do place a restriction
on the system \cite{hurst} : the magnetic
charge is restricted to take on Dirac values
({\it i.e.} $e g = n/2$ with $n$ an integer). Based on this
electromagnetic example one might similarly expect that Eq.
(\ref{2}) would place restrictions on QCD systems.
One difference between QCD and QED is that
for QED systems with field angular
momentum ({\it e.g.} the charge/monopole system or
the charge/magnetic dipole system \cite{das}) one
generally knows to a very good approximation the form
of the fields. Thus the functional form of the
field angular momentum can be determined
analytically, and the commutation
rules can be worked out explicitly. If on the other
hand the functional form of the electromagnetic
fields were not known to such a good approximation
(as is the case with the color fields of QCD)
then the angular momentum algebra might lead to broader
restrictions ({\it i.e.} ruling out a large class of
possible electromagnetic field configurations).
It would be a greater surprise if
any arbitrary form for the color fields satisfied the
angular momentum algebra. There is a
subtle point in the electromagnetic systems
with field angular momentum which needs
mentioning : the field
angular momentum {\em by itself} does not satisfy the
commutation relationships. Only the combination
of field plus particle angular momentum satisfies
Eq. (\ref{2}). Here we will find
that a similar restriction {\it might} apply
to ${\bf J}_{GB}$ -- that the pure glueball field angular
momentum must combine with some other angular
momentum in order to satisfy Eq. (\ref{2}).

\section{Standard Picture of Glueball Spin}

There are several methods for investigating glueballs.
The most rigorous method is via the numerical calculations
of lattice gauge theories. The properties of the glueballs
which are most commonly investigated in this way are
their masses, string tension, and deconfinement temperature.
In either lattice calculations or in phenomenological
models, the spin of the glueball is 
determined based on some simplifying assumptions.
For example, glueballs can be associated with
gauge invariant QCD field strength operators \cite{don}
such as $F_{\mu \nu} ^a F_a ^{\mu \nu}, F_{\mu \nu} ^a
{\tilde F}_a ^{\mu \nu}$, which are spin 0, or
$F_{\mu \lambda} ^a F_a ^{\lambda \nu}, F_{\mu \lambda} ^a
{\tilde F}_a ^{\lambda \nu}$, which are spin 2. In this
last case there is some ambiguity as to whether these
operators should be associated with one glueball of spin 2,
or two spin 0 glueballs in a d-wave state. Usually one chooses
the former case for simplicity. However, we will argue that
the angular momentum algebra may restrict glueballs to
always be spin 0, in which case the latter choice
would be required. There is also some uncertainty in the
spin assignment of glueballs in lattice studies as
pointed out in a recent lattice glueball review \cite{teper}.
Essentially for a given irreducible representation of the
lattice rotational group there are states with different
values of total angular momentum which reduce to the
same continuum state. Usually one makes the simplest choice
and associates a given continuum state with the lattice
state with the minimum value of total angular momentum. 

One can also investigate glueballs using various
phenomenological models \cite{fishb}, such as bag models
\cite{bag} or using massive constituent gluons
\cite{cs}. Both of these approaches roughly
picture the glueballs as composed of two or three
valence gluons. The glueball then has a total
spin which is taken as coming from the sum
of the orbital and spin angular momentum
of these valence gluons ({\it i.e.}
${\bf J}_{GB} = \sum ({\bf L}_i + {\bf S}_i )$ ).
This kind of picture is similar in spirit to the
quark models which had the spin of the proton
coming only from the valence quarks. Since the simple
quark models had difficulties giving the
correct proton spin structure, a similar
problem may occur with the glueball spin structure
in these simple models.

\section{Gluonic Field Angular Momentum}

Eq. (\ref{1}) can be recast in index notation as
$J^i _{GB} = \frac{1}{2} \epsilon ^{ijk} J^{jk}$
where
\begin{equation}
\label{3}
J^{jk} = \int d^3 x M^{0jk} (\vec{x})
\end{equation}
$M^{0jk}$ are some of the components of the
rank 3 tensor $M^{\alpha \mu \nu}$
\begin{equation}
\label{4}
M^{\alpha \mu \nu} = T^{\alpha \nu} x^{\mu} - T^{\alpha \mu} x^{\nu}
\rightarrow
M^{0jk} = T^{0k} x^{j} - T^{0j} x^{k}
\end{equation}
$T^{0k} , T^{0j}$ are the time-space components
of the energy-momentum tensor $T^{\mu \nu}$ of the
Yang-Mills gauge fields
\begin{equation}
\label{5}
T^{\mu \nu} = \frac{1}{4} g ^{\mu \nu}
F^{\alpha \beta a} F^a _{\alpha \beta}
- F^{\mu \alpha \; a} F^{\nu \; a} _{\alpha} \rightarrow
T^{0k} = - F^{0 i \; a} F^{k \; a} _{\; i}
\end{equation}
the Yang-Mills fields tensor components, $F^{0 i \; a} ,  F^{k i \; a}$,
are the color electric and color magnetic fields respectively.
In terms of the gauge fields these are
\begin{eqnarray}
\label{6}
F_{0i} ^a &=& \partial _0 A_i ^a -\partial _i A_0 ^a + g f^{abc}
A^b_0 A^c _i \equiv E_i ^a \nonumber \\
F_{ij} ^a &=& \partial _i A_j ^a -\partial _j A_i ^a +
g f^{abc} A^b _i A^c _j \equiv \epsilon _{ijk} B_k ^a
\end{eqnarray}
Using this string of definitions we can write down
the gluon field angular momentum of Eq. (\ref{1}) in
index form as
\begin{equation}
\label{7}
J^i _{GB} = \epsilon ^{ilm} \int d^3 x \left( F^{0n \; a}
F^{m \; a} _{\; \; n} \right) x^l
\end{equation}

\section{Angular momentum commutation relationships}

In this section we will use the expression in Eq. (\ref{7})
to calculate the commutator for the pure glueball angular momentum. 
We will not write out the integrals $\int d^3 x \int d^3 y ....$
since these can always be applied after taking the commutator
of the integrand in Eq. (\ref{7}). From Eq. (\ref{7}) we have
\begin{eqnarray}
\label{8}
[J^i _{GB} ({\bf x}) , J^j _{GB} ({\bf y})] &=& \epsilon^{ilm}
\epsilon^{jpq} x^l x^p
[(F^{0n \; a} ({\bf x}) F^{m \; a} _{\; \; n} ({\bf x})) ,
(F^{0r \; b} ({\bf y}) F^{q \; b} _{\; \; r} ({\bf y})) ]
\nonumber \\
&=& \epsilon^{ilm} \epsilon^{jpq} x^l x^p \Large(
F^{0n \; a} F^{0r \; b} [F^{m \; a} _{\; \; n} , F^{q \; b} _{\; \; r}]+
F^{0n \; a} [F^{m \; a} _{\; \; n} , F^{0r \; b} ] F^{q \; b} _{\; \; r}
\nonumber \\
&+& F^{0r \; b} [F^{0n \; a} , F^{q \; b} _{\; \; r} ] F^{m \; a} _{\; \; n} +
[F^{0n \; a} , F^{0r \; b} ] F^{m \; a} _{\; \; n} F^{q \; b} _{\; \; r}
\Large)
\end{eqnarray}
In the last line the ${\bf x}$, ${\bf y}$ dependences are
not written out explicitly.
The $\epsilon$ 's and $x^i$ 's have been pulled outside
the commutator brackets since they commute with
one another and with $F_{\mu \nu} ^a$. Standard commutator
algebra has been used to write things in terms of two-term
commutators. None of the commutators above have any
non-trivial contribution coming from the SU(3) group
structure of the fields, since the $F_{\mu \nu} ^a$ ' s are
all simply SU(3) components. Another way to see
this is to write $F_{\mu \nu} ^a$ in terms of the  matrix
field strength tensor, $F_{\mu \nu} \equiv F_{\mu \nu} ^a T^a$.
Using the standard normalization for the group generators
($Tr[T^a T^b] = \frac{1}{2} \delta ^{ab}$ where $Tr =$ Trace)
$F_{\mu \nu }^a$ can be written 
as $F_{\mu \nu} ^a =2 Tr[ F_{\mu \nu} T^a ]$.
Since the group factors are traced over in this
expression for $F_{\mu \nu} ^a$ they
only give trivial commutators in Eq. (\ref{8}).
Any possible non-trivial commutation structure in Eq. (\ref{8})
arises, as in Ref. \cite{merz}, from representing the
gauge potentials in terms of creation/annihilation operators.

In the canonical formalism, where the gauge fields
are given by creation/annihilation operators, the SU(3) gauge
fields satisfy the following commutation relationships \cite{hat}
\begin{eqnarray}
\label{10}
&[&A_i ^a ({\bf x}, t) , E_j^b ({\bf y} , t)] =
i \delta _{ij} \delta ^{ab} \delta ^3 ({\bf x-y}) \nonumber \\
&[& E_i ^a ({\bf x}, t) , E_j ^b ({\bf y} ,t) ] =
[A_{\mu} ^a ({\bf x}, t) , A_{\nu} ^b ({\bf y} , t) ] = 0
\end{eqnarray}
Expanding the chromoelectric field in the above commutators
as $E_j ^b = F_{0j} ^b = \partial _0 A_j ^b - \partial _j A_0 ^b
+g f^{bcd} A_0 ^c A_j ^d$, and using the last commutator in
Eq. (\ref{10}) one finds that only the time derivative term in the
chromoelectric field expansion gives a nontrivial commutator since
$[A_i ^a ({\bf x} , t) , \partial ^{(y)} _j A_0 ^b ({\bf y} , t)]
= \partial ^{(y)} _j [A_i ^a ({\bf x} , t) , A_0 ^b ({\bf y} , t)] = 0$
and $[A_i ^a ({\bf x} , t) ,A_0 ^c ({\bf y} , t)
A_j ^d ({\bf y} , t)] = 0$ by the
last commutator in Eq. (\ref{10}). In the first example
the superscript $(y)$ indicates that the derivative is
taken with respect to ${\bf y}$. Thus the commutation
relationships involving the chromoelectric field can be rewritten
as
\begin{eqnarray}
\label{10a}
&[&A_i ^a ({\bf x}, t) , \partial _0 A_j^b ({\bf y} , t)] =
i \delta _{ij} \delta ^{ab} \delta ^3 ({\bf x-y}) \nonumber \\
&[& \partial _0 A _i ^a ({\bf x}, t) ,
\partial _0 A_j^b ({\bf y} ,t) ] = 0
\end{eqnarray}
Except for the group index Kronecker delta these are identical
to the field commutators in an Abelian theory \cite{ms}.

When the commutator, $[F^{m \; a} _{\; \; n} ,
F^{q \; b} _{\; \; r}]$,  in Eq . (\ref{8}) is expanded
in terms of the potentials one finds commutators like
$[\partial ^m A^a _n , g f^{bef} A^{eq} A_r ^f]$ or
$[\partial ^m A^a _n , \partial ^q A^b _r]$ or
$[g f^{acd}A^{bm} A_n^c , g f^{bef} A^{eq} A_r ^f]$.
By commutator algebra these can all be reduced to
sums of commutators like $[A_n ^c , A_r ^f]$ or
$[\partial ^m A^a _n , \partial ^q A^b _r] =
\partial ^m \partial ^q [A^a _n , A^b _r]$ or
$[\partial ^m A^a _n , A^{eq}] = \partial ^m [A^a _n , A^{eq}]$
({\it i.e.} commutators of the gauge potential with itself
or with its {\it spatial} derivatives). All the these
vanish by the last commutation relationship in
Eq. (\ref{10}) thus in Eq. (\ref{8})
$[F^{m \; a} _{\; \; n} , F^{q \; b} _{\; \; r}] =0$.

The remaining three terms from Eq. (\ref{8}) involve
commutators between $F^{0i \; a}$ and itself or
between , $F^{0i \; a}$ and $F^{jk \; b}$. These
three commutators can be evaluated by expanding
$F^{0i \; a}$ and $F^{jk \; b}$ in terms of the
gauge potentials. It is easy to see that the only
non-trivial part of these three commutators comes from
the commutation of the time derivative term in
$F^{0i \; a}$ ({\it i.e.} $\partial ^0 A^{i \; a}$)
with the gauge potential ($A^{k \; b}$) or its
spatial derivative ($\partial ^j A^{k \; b}$).
Since glueballs are bound states one has the
extra physical constraint  that the gauge potentials
are time independent $\partial ^0 A^{i \; a}=0$
(at least in the glueball rest frame).
Taking this physical constraint into account
the chromoelectric part of the field strength tensor for
pure glueballs simplifies to $F^{0i \; a} = -\partial _i A_0 ^a +
g f^{abc} A^b_0 A^c _i$. (Strictly one can not set
the operator, $\partial ^0 A^{i \; a}$, equal to
zero, but this condition should be applied to the
glueball states, $\vert G \rangle$, as is done in
the following section. This Gupta-Bleuler-like
procedure could easily be applied here and would not
change the final result in Eq. (\ref{10b}) below).
From Eqs. (\ref{10}) and (\ref{10a}) one can
see that each of these terms commutes with other
gauge potential components, $A_{\mu}^a$,
or with spatial derivatives of gauge potential components.
Thus the commutators for the glueball angular momentum
operator yields
\begin{equation}
\label{10b}
[J_{GB} ^i , J_{GB} ^j] = 0
\end{equation}
The vanishing of this commutator comes
directly from the physical requirement that for bound
states the gauge field should be time-independent
($\partial ^0 A^{i \; a}=0$).

The result given in Eq. (\ref{10b})  may seem strange,
but an analogous result arises in the electromagnetic
system of a charge and monopole at rest with respect to
one another. For this system one can calculate the
field angular momentum as ${\bf J}_{EM} = e g {\bf r}/r$,
where $e, g, {\bf r}$ are the electric charge, magnetic
charge and displacement between the two charges
respectively. The commutator of the components of
${\bf J}_{EM}$ with itself gives zero since
${\bf r}/r$ commutes with itself.
Alternatively, one could proceed in the electromagnetic
case as in the color field case by writing
${\bf J}_{EM}$ out as in Eq. (\ref{1}) in terms of
the normal electric and magnetic fields. The commutator
of the components of ${\bf J}_{EM}$ among themselves
could then be calculated using the electromagnetic
version of the gauge potential commutators given by
Eqs. (\ref{10}) (\ref{10a}) with the color indices
dropped. The only non-trivial terms would come
from commutators between $\partial ^0 A^i$ and
$A^j$ or $\partial ^j A^k$. However, in the frame
where the electric and magnetic charges are at rest,
the gauge fields are time-independent so that
$\partial ^0 A^i = 0$, and one again finds that the
commutator for the electromagnetic field angular momentum
is zero. In the color field case this latter method
is the only viable one, since for the color interaction
an explicit form for ${\bf E}^a$ and ${\bf B}^a$ is
not known so that an explicit calculation of
${\bf J}_{QCD}$ is not possible. For the electromagnetic
system ${\bf E}$ and ${\bf B}$ are Coulomb fields, and
so ${\bf J}_{EM}$ can be calculated directly.

There are several ways of looking at this apparent
conflict between Eqs. (\ref{2}) and (\ref{10b}).
\begin{itemize}
\item
First, Eqs. (\ref{2}) and (\ref{10b}) are
consistent if glueballs  are restricted to be
spin 0 (${\bf J}_{GB} =0$) bound states. This is
not an unreasonable restriction since the identification
of the spin of a non-zero spin glueball is usually
based on some simplifying assumptions. As mentioned
previously spin 2 glueballs are usually connected with
QCD operators like $F_{\mu \lambda} ^a F_a ^{\lambda \nu}$ or
$F_{\mu \lambda} ^a {\tilde F}_a ^{\lambda \nu}$. However,
one could instead associate these objects with two spin 0
glueballs in a d-wave orbital angular momentum state. The
former choice is a reasonable, simplifying assumption, but
may not be correct in light of the restriction coming from
$[J_{GB} ^i , J_{GB} ^j] =0$.
\item
A second explanation of the zero commutator is that glueballs
do not occur as pure glue bound states, but always
have some admixture of quarks. If the glueball states
are mixed with ${\bar q} q$ states, these would
contribute to the overall spin of the bound state.
To take into account the contribution coming from
the quarks, one would add the total quark
angular momentum operator \cite{ji}
\begin{equation}
\label{11}
{\vec J}_Q = \int d^3 x \left[ {1 \over 2} {\bar \psi} {\vec \gamma}
\gamma _5 \psi + \psi ^{\dag} ({\vec x} \times (-i {\vec D})) \psi \right]
\end{equation}
to the pure QCD term of Eq. (\ref{1}) so that
${\vec J} _{GB} \rightarrow {\vec J} _{GB}+ {\vec J}_Q$.
(The two terms in Eq. (\ref{11}) are associated with the
spin and orbital angular momentum of the quarks respectively).
This total angular momentum might then satisfy Eq. (\ref{2})
with the proper non-zero term on the right hand side. Such mixing of
quarks with glueballs is thought to be suppressed via the
$1/N_c$ expansion approach \cite{don}. However, the restriction
coming from the angular momentum commutators as discussed
above, might be an indication that
pure glue states can not exist -- that there will always be
some substantial admixture of quarks in order to have
the correct angular momentum commutators. This postulated
mixing might explain the difficulty in experimentally
distinguishing an object as a pure glueball. This
explanation for Eq. (\ref{10b}) is also
reminiscent of the electromagnetic charge/monopole
system, where it is only the {\em sum} of the field
angular momentum plus the other contributions which
satisfies Eq. (\ref{2}).
\item
Finally, recent work \cite{dva} \cite{ji2} has shown that
the choice of the Lorentz frame in which the angular
momentum of a system is evaluated can have an impact
on the interpretation of the angular momentum. Thus,
the peculiar result of Eq. (\ref{10b}) may be connected
with the choice of Lorentz frame in which the glueball
angular momentum is evaluated (up to this point
we have implicitly assumed that $J^i _{GB}$
is evaluated in the rest frame of the glueball).
In Ref. \cite{dva} an investigation of the $(\frac{1}{2} ,
\frac{1}{2})$ representation of the Lorentz group,
showed that it is not possible, in a general
Lorentz frame, to split the
$(\frac{1}{2}, \frac{1}{2})$ space into spin 1 and spin 0
sectors. Two special cases were found in which the spin 1 / spin 0
decomposition was possible : (a) in the rest frame
(b) in the helicity frame where the quantization axis for spin
projections is aligned with the boost direction. 
In Ref. \cite{ji2} a related result was found in
a somewhat different context. Ref. \cite{ji2} examined
the break up of the total nucleon spin into terms
for the quark spin and orbital angular momentum,
and gluon spin and orbital angular momentum. Under
a general Lorentz transformation it was found that
these individual terms did not transform properly, thus
making their interpretation frame dependent. Ref. \cite{ji2}
also found that the two special frames listed above
({\it i.e.} the rest frame and the helicity frame) were picked out
as being particularly useful in studying the spin structure of the
nucleon. Although the central concerns of each of these
works differ from each other, they both point out the
importance that choice of the Lorentz frame
makes in the study of the angular momentum.
This may imply that the difficulty in formulating a
clear picture of the spin structure of the glueball
may rest in our implicit choice of the rest frame.
\end{itemize}

There are certain phenomenological bag models of glueballs
\cite{bag1} where the assumption that the color fields
are time independent is not true. In these models,
to lowest order, the chromoelectric and chromomagnetic fields are
analogous to the ordinary electric and magnetic fields
inside a resonant cavity so that the chromoelectric and
chromomagnetic fields, and the gauge potentials
have oscillatory time-dependences. However the chromoelectric
and chromomagnetic fields in these models are $90 ^o$
out of phase with one another. Therefore at some particular
time the chromoelectric field will be at a maximum while the
chromomagnetic field is zero, and at some later time
the chromomagnetic field will be at a maximum while the
chromoelectric field vanishes. At these times, when either
${\bf E}^a$ or ${\bf B}^a=0$,  ${\bf J} _{GB} =0$
from Eq. (\ref{1}). Since ${\bf J} _{GB}$ is conserved
({\it i.e.} is time independent) this implies that
${\bf J} _{GB}=0$ at any time. This vanishing of
${\bf J} _{GB}$ was one of the possible resolutions
that we arrived at by applying the angular momentum
commutators to ${\bf J} _{GB}$ and assuming that the
fields were time independent. Thus, even in the phenomenological
bag models where the time independence assumption does not hold,
we still find the restriction ${\bf J} _{GB} =0$.
One can also present a more general argument to
support the time independence of the fields.
Since ${\bf J} _{GB}$ is conserved $\rightarrow$
$d{\bf J} _{GB}/ dt =0$. From Eq. (\ref{1})
this implies that
\begin{equation}
\label{last}
\int d^3 x \left[
{\bf x} \times \left( {d {\bf E}^a \over dt} \times {\bf B}^a
+{\bf E}^a \times {d {\bf B}^a \over dt} \right) \right] = 0
\end{equation}
(${\bf x}$ is not differentiated since it is the
time independent position vector from the origin
to some particular part of the momentum density,
${\bf E}^a \times {\bf B}^a$).
There are two simple ways in which Eq. (\ref{last})
can be satisfied. First, if $d{\bf E}^a / dt =
d{\bf B}^a / dt =0$. This implies that the fields
are time independent which was the assumption in our
analysis. Second, if ${\bf E}^a = \pm {\bf B}^a$
then Eq. (\ref{last}) is satisfied. However, in this
case, even though the fields do not need to be
time independent, one finds that ${\bf J}_{GB} =0$
from Eq. (\ref{1}). There is apparently no simple
way to have {\it both} ${\bf J}_{GB} \ne 0$
{\it and} $d{\bf E}^a / dt \ne 0$, $d{\bf B}^a / dt \ne 0$.

\section{Expectation of ${{\bf J}_{GB}}$}

In this section we calculate the expectation value
of $J_{GB}^i$ with respect to a glueball state,
$ \vert G \rangle$, under certain simplifying assumptions.
From Eq. (\ref{7}) the expectation value is given by
\begin{equation}
\label{12}
\langle G \vert J^i _{GB} \vert G \rangle = \epsilon ^{ilm}
\int d^3 x \langle G \vert\left( F^{0n \; a} ({\bf x})
F^{m \; a} _{\; \; n} ({\bf x}) \right) \vert G \rangle x^l
\end{equation}
The glueball state is taken to be normalized so that
$\langle G \vert G \rangle =1$. Since the field angular
momentum operator in Eq. (\ref{7}) is gauge invariant
we are free to choose a gauge without effecting the
physically measurable quantities like the spin of the
glueball. We choose the temporal gauge where $A_0 ^a =0$.
In terms of the operator $A_0 ^a$ we implement
this gauge choice using a Gupta-Bleuler-like approach by
making it  a condition on the glueball state --
$\langle G \vert A_0 ^a \vert G \rangle =0$. $A_0 ^a$ can be split
into positive and negative frequency parts : $A_0 ^a =
A_0 ^{a(+)} + A_0 ^{a(-)}$. The positive frequency part can be
written as a sum of annihilation operators, while the
negative frequency part can be written as a sum of
creation operators. Thus the {\em gauge} condition
$\langle G \vert A_0 ^a \vert G \rangle =0$ becomes
$A_0 ^{a(+)} \vert G \rangle =0$ and
$\langle G \vert A_0 ^{a(-)} =0$.
Next, since the glueball is
a bound state, we have the {\em physical} (as opposed to gauge)
restriction that $\partial _0 A_i ^a = 0$ which we will
also implement as a condition on the glueball state --
$\langle G \vert \partial _0 A_i ^a \vert G \rangle = 0$.
This can also be written in terms of the positive and negative
frequency parts as $\partial _0 A_i ^{a(+)} \vert G \rangle = 0$
and $\langle G \vert \partial _0 A_i ^{a(-)} = 0$. Expanding
$F^{0n \; a}$ in term of the gauge potentials we can write
out a portion of the right hand side of Eq. (\ref{12}) as
\begin{equation}
\label{13}
\left( \partial ^0 A ^{n a} ({\bf y}) - \partial ^n A ^{0 a} ({\bf y})
+ g f^{abc} A^{0b} ({\bf y}) A^{nc} ({\bf y}) \right) F^{m \; a} _n ({\bf x})
\vert G \rangle
\end{equation}
By using the commutation rules for the non-Abelian
gauge potentials given by Eq. (\ref{10a}) we will move
the various terms coming from $F^{0n \; a}$ through
$F^{m \; a} _n$ and then apply either the gauge
condition $A_0 ^{a(+)} \vert G \rangle =0$ or the physical
condition $\partial _0 A_i ^{a(+)} \vert G \rangle = 0$.
In order to apply the commutators of Eq. (\ref{10a}) we have
changed $F^{0n \; a} ({\bf x})$ to $F^{0n \; a} ({\bf y})$.
In the end we will let ${\bf y} \rightarrow {\bf x}$.
In Eq. (\ref{13}) $F^{0n \; a}$, expanded in terms
of the $A^{\mu a}$'s, has both positive and negative frequency
parts, but the negative frequency parts will vanish directly
when $\langle G \vert$ is applied to the left side
of Eq. (\ref{13}). Thus as we pull $F^{0n \; a}$
through $F^{m \; a} _n$ we are only dealing with the
positive frequency parts even though the superscript
$(+)$ will not be written out explicitly.
The last term in $F^{0n \; a}$ ($g f^{abc} A^{0b} ({\bf y})
A^{nc} ({\bf y})$) can be trivially commuted past
$F^{m \; a} _n ({\bf x})$ since both  $A^{0b} ({\bf y})$ and
$A^{nc} ({\bf y})$ commute with other gauge fields components or
their {\it spatial} derivatives by the last commutator in
Eq. (\ref{10}). After this term is commuted through
$F^{m \; a} _n ({\bf x})$ one has
\begin{equation}
F^{m \; a} _n ({\bf x}) [A^{0b} ({\bf y}) A^{nc} ({\bf y})] \vert
G \rangle  \rightarrow  F^{m \; a} _n ({\bf x})
[A^{nc} ({\bf y}) A^{0b} ({\bf y})] \vert G \rangle =0
\end{equation}
where the gauge condition $A_0 ^a \vert G \rangle =0$
was applied. The second term from Eq. (\ref{13}),
$\partial ^n A ^{0 a} ({\bf y})$, can also be commuted
past $F^{m \; a} _n ({\bf x})$ using the last commutator
in Eq. (\ref{10}) since only a {\it spatial} derivative is
involved. This yields
\begin{equation}
F^{m \; a} _n ({\bf x}) [\partial ^n A ^{0 a} ({\bf y})] \vert
G \rangle  = 0
\end{equation}
where we have taken the gauge condition $A_0 ^a \vert G \rangle =0$
to imply $\partial ^n (A_0 ^a \vert G \rangle) =0$. Finally
we commute the term, $\partial ^0 A ^{n a} ({\bf y})$
through $F^{m \; a} _n ({\bf x})$ by expanding
$F^{m \; a} _n ({\bf x})$ as
\begin{equation}
F^{m \; a} _n ({\bf x}) = \partial ^m A_n ^a ({\bf x})
- \partial ^n A_m ^a ({\bf x})  + g f^{ade} A^{md} ({\bf x})
A_n ^e ({\bf x})
\end{equation}
From Eq. (\ref{10}) $[\partial ^0 A ^{n a} ({\bf y})  ,
\partial ^n A_m ^a ({\bf x})] = -i \delta _{mn}
\partial ^n [ \delta ^3 ({\bf y-x})] = 0$ since
$m \ne n$. Thus $\partial ^0 A ^{n a} ({\bf y})$
commutes with the $\partial ^n A_m ^a ({\bf x})$ term
from $F^{m \; a} _n ({\bf x})$. Also
$\partial ^0 A ^{n a} ({\bf y})$ will commute
past both fields operators in the term
$f^{ade} A^{md} ({\bf x}) A_n ^e ({\bf x})$.
This is due to $a \ne d \ne e$ from the antisymmetry
of $f^{ade}$, and the Kronecker delta in the group
indices in the commutator of Eq. (\ref{10})
({\it i.e.} $[\partial ^0 A ^{n a} ({\bf y})  ,
A_n ^e ({\bf x})] = -i \delta _{nn} \delta ^{ae}
\delta ^3 ({\bf y-x}) = 0$ since $a \ne e$). Thus
\begin{eqnarray}
\partial ^0 A ^{n a} &(&{\bf y}) \partial ^n A_m ^a ({\bf x})
\vert G \rangle \rightarrow
\partial ^n A_m ^a ({\bf x}) \left( \partial ^0 A ^{n a} ({\bf y}) 
\vert G \rangle \right) =0 \nonumber \\
\partial ^0 A ^{n a} &(&{\bf y}) f^{ade} A^{md} ({\bf x})
A_n ^e ({\bf x}) \vert G \rangle \rightarrow
f^{ade} A^{md} ({\bf x}) A_n ^e ({\bf x})
\left( \partial ^0 A ^{n a} ({\bf y})
\vert G \rangle \right) = 0
\end{eqnarray}
where we have used the physical condition
$\partial _0 A_i ^a \vert G \rangle = 0$. Finally we examine the
last term, $\partial ^0 A^{na} ({\bf y}) \partial ^m A_n ^a ({\bf x}) 
\vert G \rangle$. Using the first commutator in Eq. (\ref{10a}) 
(so $\partial ^0 A^{na} ({\bf y}) \partial ^m A_n ^a ({\bf x}) =
\partial ^m A_n ^a ({\bf x}) \partial ^0 A^{na} ({\bf y}) -
i \partial _{(x)} ^m (\delta ^3 ({\bf x-y}))$ where the subscript
$(x)$ means that the derivative is taken with respect to ${\bf x}$
rather than ${\bf y}$). Using this gives
\begin{eqnarray}
\label{15}
\partial ^0 A^{na} ({\bf y}) \partial ^m A_n ^a ({\bf x})
\vert G \rangle &=& \left[ \partial ^m A_n ^a ({\bf x})
\partial ^0 A^{na} ({\bf y}) -
i \partial _{(x)} ^m (\delta ^3 ({\bf x-y})) \right] \vert G \rangle
\nonumber \\
&=& - i \partial _{(x)} ^m (\delta ^3 ({\bf x-y})) \vert G \rangle
\end{eqnarray}
where the physical condition $\partial ^0 A^{na} \vert G \rangle = 0$
was used. Using Eq. (\ref{15}) one obtains
\begin{equation}
\label{16}
\langle G \vert J_{GB}^i \vert G \rangle = - \epsilon^{ilm} \int d^3 x
\langle G \vert x^l (i \partial _{(x)} ^m [\delta^3 ({\bf x-y})]) 
\vert G \rangle = - \epsilon^{ilm} \int d^3 x \left(
i \partial _{(x)} ^m [\delta^3 ({\bf x-y})] \right) x^l 
\end{equation}
where $\langle G \vert G \rangle =1$ was used. Integrating Eq.
(\ref{16}) by parts gives
\begin{equation}
\label{17}
- i \epsilon ^{ilm} [x^l \delta ^3 ({\bf x-y})]
{\Big \vert} _{-\infty} ^{+\infty}
+ \int \delta ^3 ({\bf x-y}) (\partial _{(x)} ^m x^l ) d^3 x
\end{equation}
The first term above is zero because of the $\delta$ function.
The integral in the second term evaluates to the Kronecker delta,
$\delta ^{ml}$. Thus the expectation of $J^i _{GB}$ becomes
\begin{equation}
\label{18}
\langle G \vert J_{GB}^i \vert G \rangle = i \epsilon ^{ilm} \delta ^{ml}
= i \epsilon ^{ill} = 0
\end{equation}
using the properties of the $\epsilon$'s. Thus by imposing
the temporal {\em gauge} condition
($A_0 ^{a(+)} \vert G \rangle =0$ ; $\langle G \vert A_0 ^{a(-)}=0$) and the
{\em physical} condition that for a bound state the gauge fields
should be time independent ($\partial ^0 A^{ia(+)} \vert G \rangle =0$ ;
$\langle G \vert \partial ^0 A^{ia(-)} =0$)
it is found that $\langle G \vert J_{GB}^i \vert G \rangle = 0$.  

\section{Conclusions}

By examining the commutation relationships ($[J_{GB}^i , J_{GB}^j]$) 
and the expectation value ($\langle G \vert J_{GB}^i \vert G \rangle$) 
of the glueball angular momentum operator it was found that both
of these quantities were zero. The main assumption in both of these 
results was that the gauge fields were time-independent 
($\partial ^0 A^{ia} =0$), since the glueball is a bound state.
The non-Abelian nature of the gauge fields played
little role in the commutation relationships of $J_{GB}^i$, since
the group indices are traced over in the expression of
$J_{GB}^i$. Thus one would expect that a similar result
($[J_{EM}^i , J_{EM}^j]=0$ where $J_{EM} ^i$ is the
electromagnetic field angular momentum) should hold
for Abelian gauge fields ($A_{\mu}$) under similar conditions
($\partial ^0 A^i =0$). For the two Abelian examples where 
this has been worked out explicitly (the monopole/charge
system and the magnetic dipole/charge system \cite{das}) it 
is in fact found that the field angular momentum, {\it by itself},
does not satisfy the angular momentum commutation relationships.
In light of these Abelian examples, and the similarity between
$J_{GB}^i$ and $J_{EM}^i$ the results for the glueball angular
momentum operator are not so surprising.

We have suggested several possible resolutions/explanations for
these results for $J_{GB}^i$.
\begin{itemize}
\item
First, the fact that both
$[J_{GB} ^i , J_{GB} ^j]$ and $\langle G \vert J_{GB}^i \vert G \rangle$
are zero may imply that pure glueballs can only be spin 0.
\item
Second, the above results may indicate
that glueballs will always have some significant admixture
of quarks, which will contribute to the spin structure
of the glueball.
\item
Finally, one needs to consider the effect that the choice
of Lorentz frame makes in the study and interpretation of
the angular momentum of the glueball system. 
\end{itemize}

\section{Acknowledgments} I would like to thank Xiangdong Ji
who suggested this as a problem. I also want to thank Gerardo Munoz
and Dharam Ahluwalia for discussions and comments on this paper.

\end{document}